## Mesoscopic twin boundaries in epitaxial Ni-Mn-Ga films

Anja Backen[1,2,a,b] , Sandra Kauffmann-Weiss[1,2,3], Christian Behler[1,3], Anett Diestel[1,2], Robert Niemann[1,3], Alexander Kauffmann[1,2], Jens Freudenberger[1,4], Ludwig Schultz[1,2] and Sebastian Fähler[1,3]

[1] IFW Dresden, P.O. Box 270116, 01171 Dresden, Germany

[2] Dresden University of Technology, Department of Mechanical Engineering, Institute for Materials Science, 01062 Dresden, Germany

[3] Technische Universität Chemnitz, Faculty of Natural Sciences, Institute of Physics, 09107 Chemnitz, Germany

[4] Technische Universität Bergakademie Freiberg, Institut für Werkstoffwissenschaft, Akademiestr. 6, 09599 Freiberg, Germany

### Abstract

Twin boundaries play an essential role in the use of magnetic shape memory alloy Ni-Mn-Ga as active material. Only if twin boundaries can be moved by an external magnetic field, high strain values of up to 10 % can be obtained. Therefore, understanding the observed twin microstructure of thin films is crucial for future application. We exemplarily present two different microstructural pattern, type X and type Y microstructure, using the example of two Ni-Mn-Ga films with similar film thickness and composition. The analysis of microstructure and structure of the thin films shows that both, type X and type Y pattern, are formed by mesoscopic 14M twin boundaries. The mesoscopic 14M twin boundaries are either tilted from the substrate surface in case of type X pattern or they are perpendicular to the surface for type Y pattern. Based on a recently proposed scenario for the nucleation of martensite in austenite, we can trace back the difference in twin boundary orientation to differently oriented martensitic nuclei.

### Introduction

The integration of active materials in microsystems requires the combination of high achievable strains with high cycling frequencies. The magnetic shape memory (MSM) alloy Ni-Mn-Ga shows high prospects for this particular field of application. In Ni-Mn-Ga, an external magnetic field can induce a macroscopic strain of up to 10 % [1] which exceeds the maximum strain for conventional piezoelectric materials by an order of magnitude [2]. Furthermore, cycling

[a] Corresponding author. Email: anja.backen@neel.cnrs.fr

[b] Current address: Institut Néel (CNRS & UJF), 25 Rue des Martyrs, BP166, 38042 Grenoble Cedex 9, France

frequency can reach values of 250 Hz [3] up to 2 kHz [4] in pulsed magnetic fields. The underlying mechanism of the magnetic shape memory effect is the magnetically induced reorientation (MIR) of martensitic variants. When applying an external magnetic field, the alignment of the magnetic easy axis along the external magnetic field is favored due to the high magnetocrystalline anisotropy of the martensite phase [5, 6]. This reorientation of variants is realized by introducing (101) twin boundaries between different martensite orientations and eventually moving them through the crystal. One crucial requirement for magnetic field induced twin boundary motion is that the twinning stress of the (101) twin boundaries does not exceed the maximum stress applicable by a magnetic field. This maximum is in the order of 2 MPa [7].

The twinning stress strongly depends on the martensitic phase of Ni-Mn-Ga. In the tetragonal non-modulated (NM) martensite, a mechanical stress value of several tens of MPa is necessary to move (101) twin boundaries. In contrast to NM martensite, twin boundary motion can be induced magnetically in the orthorhombic 10-layer (10M) [8] and 14-layer (14M) [1] modulated martensites, respectively, since these phases show a twinning stress well below the critical value of 2 MPa.

In addition, one has to distinguish type I and type II twin boundaries. These different types result from the underlying symmetry operations that are performed in order to create the twin boundary. Type I twin boundaries can be described by mirroring the martensitic unit cell at a (101) plane. In case of type II twin boundaries, the martensitic unit cell is rotated around an axis which lies within the (101) plane. In case of the tetragonal NM martensite, these two symmetry operations are equivalent. The resulting twin boundaries are called compound boundaries [8]. However in case of the monoclinic modulated phases, mirroring and rotation operations do not result in identical states. Therefore, the twin boundaries have different structures which leads to the observation of significantly different values for the twinning stress. Straka et al. [8] have shown, that in case of 10M martensite, the twinning stress of type I twin boundaries is about 1 MPa, whereas the twinning stress of type II twin boundaries can be as low as 0.05 MPa. In conclusion, the structure of martensite and the type of twin boundaries are crucial parameters that decide if twin boundary motion can be induced magnetically.

In bulk Ni-Mn-Ga, all six orientations of (101) twin boundaries are equivalent. This changes when changing the sample dimension to thin Ni-Mn-Ga films due to the constraint of the rigid substrate. One has to distinguish between two different orientations of twin boundaries with respect to the substrate surface. The twin boundaries are inclined to the substrate surface by either 45° or by 90°. The different orientations are sketched in Figure 1. The traces of the twin boundaries form characteristic microstructural pattern at the film surface. We assign the observed microstructure to either type X or type Y pattern respectively. Yang et al. [9] recently described the pattern and concluded that they result from twinned NM martensite phases. Our comprehensive analysis of the microstructure and the structure in combination with a nucleation model of martensite [10] suggests that both microstructures result from twinned 14M martensite.

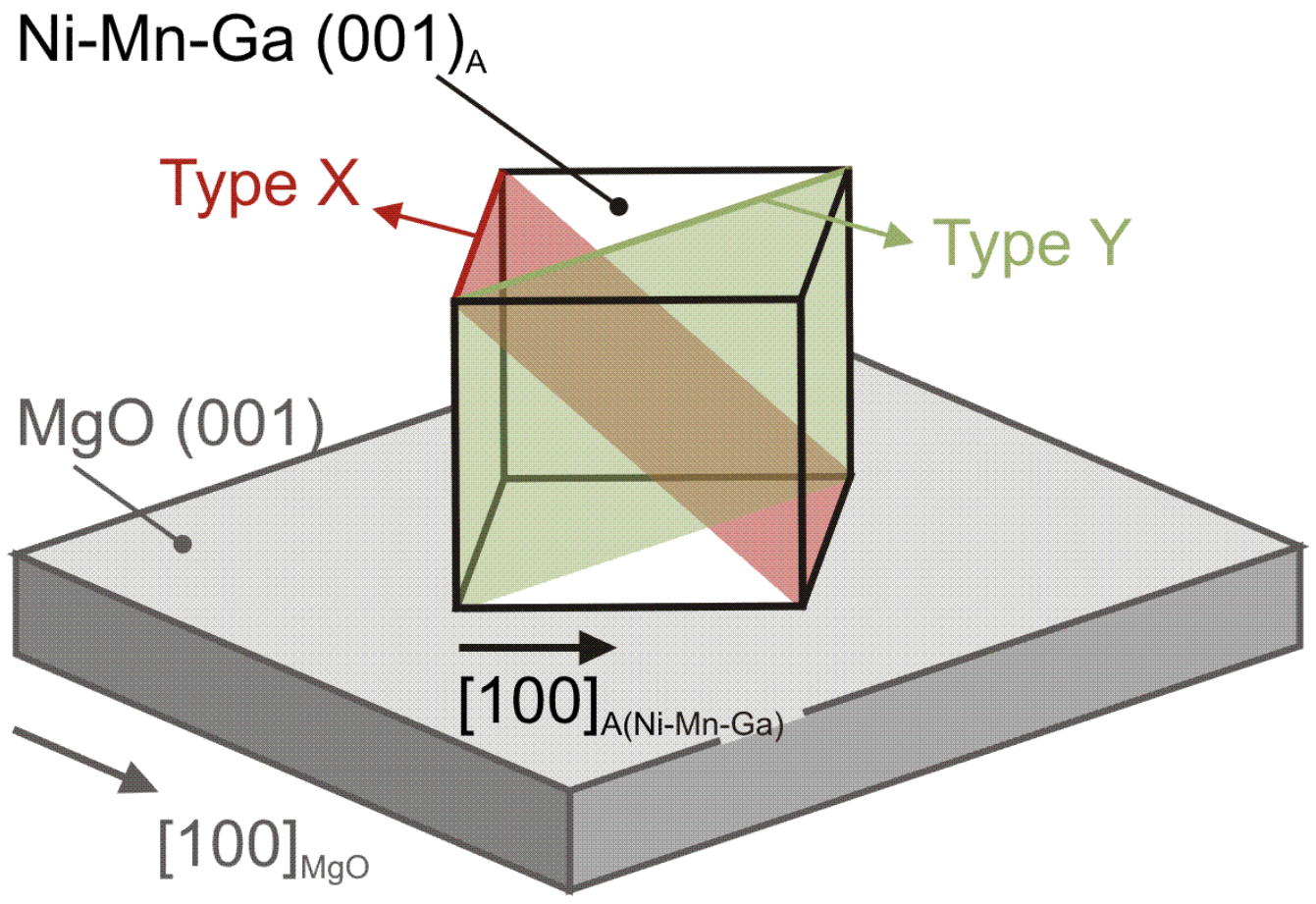

Figure 1: In thin Ni-Mn-Ga films, two non-equivalent orientations of twin boundaries can be distinguished: twin boundaries being inclined to the film surface by either 45° or by 90°, respectively. The traces of twin boundaries at the film surface form characteristic microstructural pattern that are referred to as type X and type Y pattern, respectively.

## Experimental Procedure

We prepared epitaxial Ni-Mn-Ga films by magnetron sputtering from an alloyed $Ni_{44}Mn_{32}Ga_{24}$ target. The single crystalline MgO(001) substrate was first covered with an epitaxial Cr buffer. Cr and Ni-Mn-Ga was deposited at a substrate temperature of 400 °C in order to enable epitaxial growth and chemical ordering. The epitaxial relationship between MgO substrate, Cr buffer and Ni-Mn-Ga film is MgO(001)[100]‖Cr(001)[110]‖Ni-Mn-Ga(001)[110] [11]. The nominal film thickness of Ni-Mn-Ga is 1.5 µm and 2 µm, respectively.

The film microstructure was probed by Scanning Electron Microscopy (SEM) with a LEO 1530 Gemini (Zeiss). We used backscatter electron (BSE) contrast while the sample was tilted by 7° around the substrate normal. Energy dispersive X-ray Spectroscopy (EDX) on a Phillips SEM with integrated EDAX system was used to determine the film compositions of $Ni_{48.6}Mn_{32}Ga_{19.4}$ and $Ni_{48.4}Mn_{32.8}Ga_{18.8}$ for the 2 µm and the 1.5 µm thick Ni-Mn-Ga films, respectively. The microstructure and the surface profiles of the films were determined by Atomic Force Microscopy (AFM) with DI Dimension 3100 in tapping mode. The data analysis was performed using the free software WSxM [12].

The film structure was analyzed by X-Ray diffraction in a Bruker D8 X-Ray Diffractometer. To probe all martensitic variants, the X-ray diffraction measurements were performed with □-offsets of 1° to 10°. In order to investigate the course of the twin boundaries through the film thickness, cross-section samples were prepared by in-situ lift-off technique in a dual beam Focused Ion Beam (FIB) system (FEI Helios 600i) capable of low-kV $Ga^+$ ion beam milling and investigated by Transmission Electron Microscopy (TEM) using a FEI Tecnai T20, equipped with a $LaB_6$ cathode, operating at acceleration voltages of 200 kV.

## Results

### Microstructural pattern and corresponding twin boundaries

The SEM micrographs of two films with similar film thicknesses of 2 µm and 1.5 µm are shown in Figure 2 a and Figure 2 c & d, respectively. The films have the same composition which was determined by EDX. The micrographs show traces of twin boundaries that are oriented differently with respect to the austenitic Ni-Mn-Ga unit cell and consequently form different microstructural pattern. We assign the observed microstructural pattern to type X and type Y pattern. Type X pattern is governed by traces of twin boundaries that are approximately parallel to the $[010]_A$ direction of the austenitic Ni-Mn-Ga unit cell. A closer look reveals the existence of different laminates with total width of up to 5 µm. These laminates are rotated by ±6° from the $[010]_A$ direction and thus form a rhombohedral structure with an angle of about 12° between two laminates (Figure 2 a). The FIB cross section (CS) was performed perpendicular to the observed traces at the film surface. An exemplary cutting orientation is marked in Figure 2a. The cross section (Figure 2 b) reveals that the type X pattern is formed by twin boundaries that are rotated about 45° with respect to the film surface.

A different microstructural pattern, that we refer to as type Y pattern, was observed at the surface of a 1.5 µm thick Ni-Mn-Ga film (Figure 2 c & d). In this case, the traces of twin boundaries at the film surface are parallel to the $[110]_A$ and $[-110]_A$ directions of austenite. The observed type Y pattern resembles a marquetry pattern. It is composed of straight lines that form laminates with lengths of several hundred micrometers, the total width of these laminates does not exceed 10 µm. Furthermore the width of the individual variants within the laminates seems to increase when approaching the middle of the laminate. The individual laminates cross under an angle of about 90°. In between, we either observe type Y pattern with irregular variant width or type X pattern. The FIB cross cut (Figure 2e) shows that type Y pattern results from twin boundaries that are almost perpendicular to the film surface.

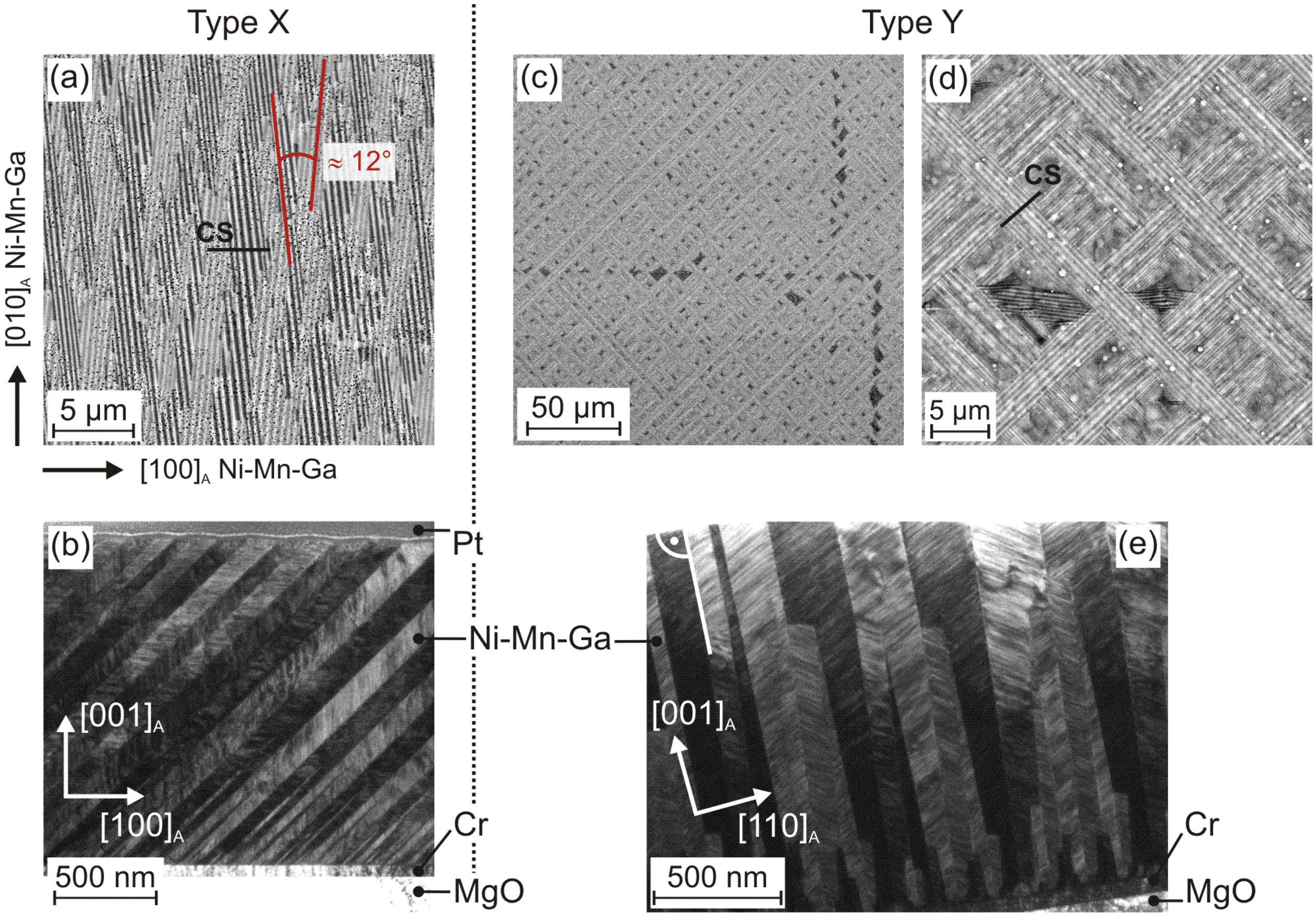

Figure 2: (a) Type X pattern is visible in SEM micrograph of 2 µm thick film and shows traces of twin boundaries that are approximately parallel to $[010]_A$ direction of Ni-Mn-Ga austenite. Individual laminates form an angle of about 12°. (b) The FIB cross section (CS) perpendicular to the traces at the surface reveals that type X pattern is a result of twin boundaries that are rotated by 45° with respect to the film surface. (c) Type Y pattern is observed in SEM micrograph of 1.5 µm thick film and reveals traces of twin boundaries that are aligned along $[110]_A$ and $[-110]_A$ direction of austenite, respectively. These twin boundaries form a marquetry pattern on the film surface (d) with type X pattern in between. The FIB cross section (e) shows that type Y pattern is composed of twin boundaries that are almost perpendicular to the film surface.

## Surface topography of type X pattern

Further differences between type X and type Y pattern can be observed in the height profiles. Figure 3a shows an AFM micrograph of the 1.5 µm thick film where both microstructural pattern are visible next to each other. The corresponding height profile of type X (P1 in Figure 3b) is very regular. Each twin boundary is represented by a minimum or a maximum in the curve. This height profile has been observed in previous studies [13, 14] and can be assigned to a twinned laminate of 14M martensite. It could be shown that the short and magnetic easy $c_{14M}$-axis alternates between in-plane and out-of-plane orientation. Furthermore, Diestel et al. [15, 16] used this regular height profile at the film surface for the analysis of the twinning periodicity of Ni-Mn-Ga films. The twinning periodicity represents the width of two adjacent variants. They could show that the twinning periodicity only depends on the film thickness. Thus in case of type X pattern, one can conclude that the width of adjacent variants at the film surface is constant for a given film thickness.

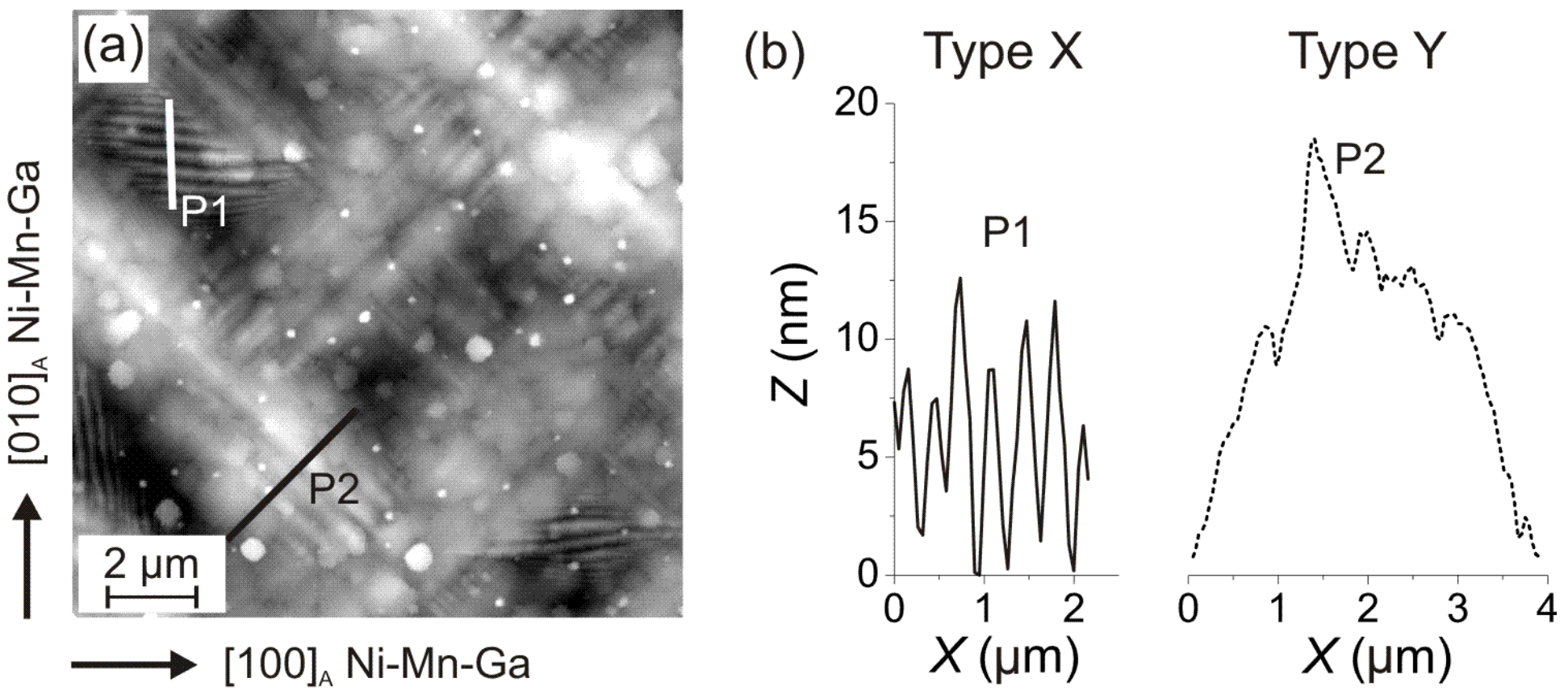

Figure 3: (a) AFM micrograph of 1.5 µm thick film shows type X and Y pattern next to each other. (b) The corresponding height profiles show a regular profile for type X (P1) and an irregular profile for type Y (P2) pattern.

**Surface topography of type Y pattern: Quantitative analysis of variant width using Hough transform**

In contrast to the regular profile of type X (P1 in Figure 3b), the type Y pattern (P2 in Figure 3b) reveals an irregular height profile. The individual twin boundaries cannot be resolved in AFM micrograph. Furthermore it is evident, that the height of the laminates forming a type Y pattern increases from the border to the middle of the laminate. Due to the irregularity, the height profile cannot be used to determine the width of individual martensitic variants. For further evaluation of the variant width, we quantitatively analyzed high resolution SEM micrographs (Figure 4 a & d). The location of the twin boundary is determined by means of an edge detection filter. Afterwards the distance between the individual twin boundaries can be analyzed using Hough transform. For this purpose, a fixed origin is chosen, in our case the origin is located in the middle of the SEM micrograph (Figure 4 a). Each line in the SEM micrograph can now be described with a certain distance *d* from the origin and an angle α that is sketched in Figure 4a. Straight parallel lines in the SEM micrograph are represented by intensity maxima in Hough space (Figure 4 b). The intensity maxima can be observed at constant angle but with varying distance from the origin. Since the twin boundary traces that form a type Y pattern represent straight lines that run almost parallel to the $[-110]_A$ direction of austenite, the cut through Hough space in the range of $42.25° < \alpha < 42.75°$ (Figure 4c) shows each twin boundary as a maximum in intensity. The distance between the maxima gives the width of individual variants.

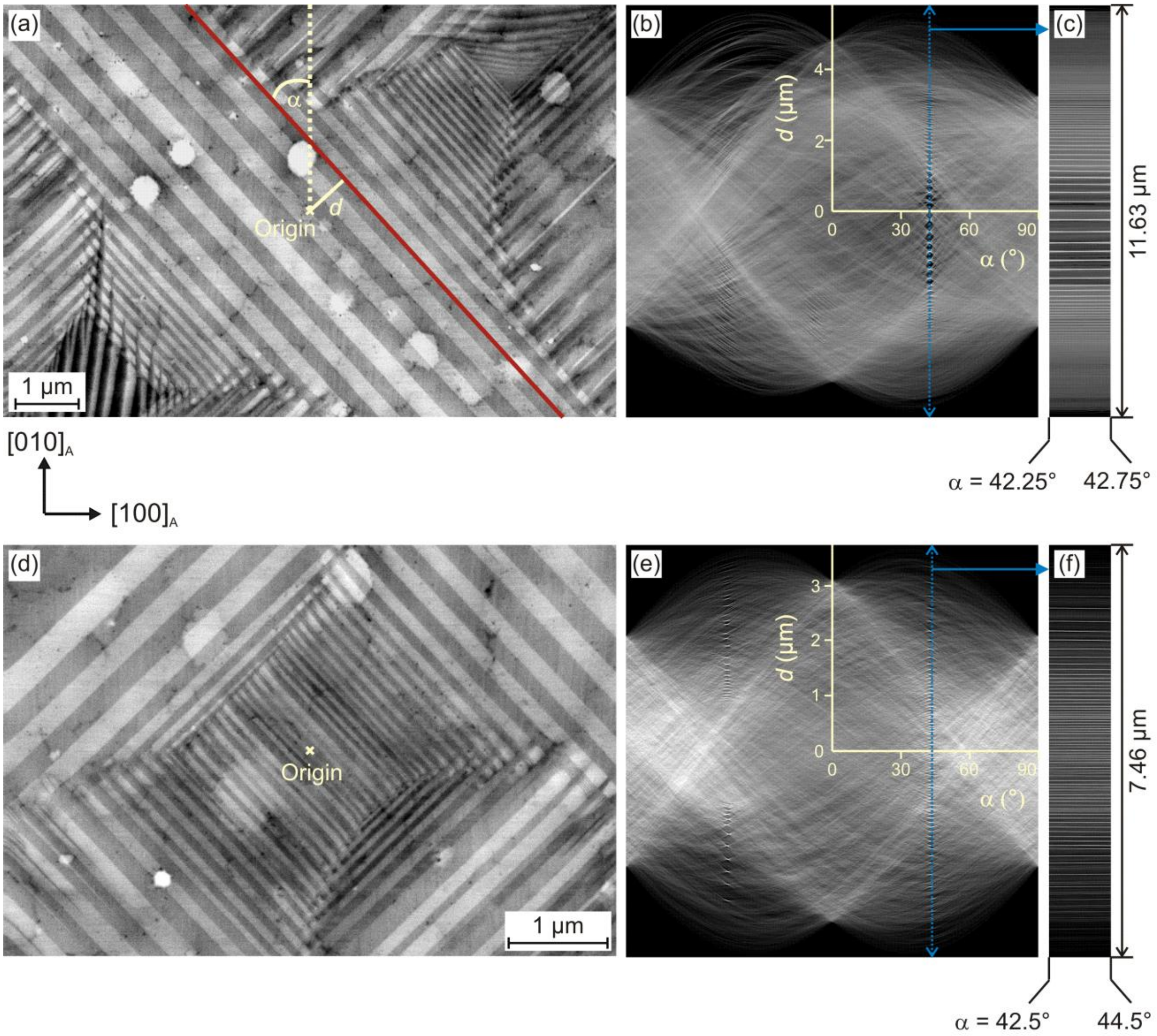

Figure 4: High resolution SEM micrographs (a,d) reveal different regions of type Y pattern on the surface of 1.5 µm thin Ni-Mn-Ga film. Twin boundaries (red line in (a)) can be described by the distance *d* from the origin and the angle α (a) and are represented by intensity maxima at constant angle α in the corresponding Hough space (b,e) which is magnified in c and f.

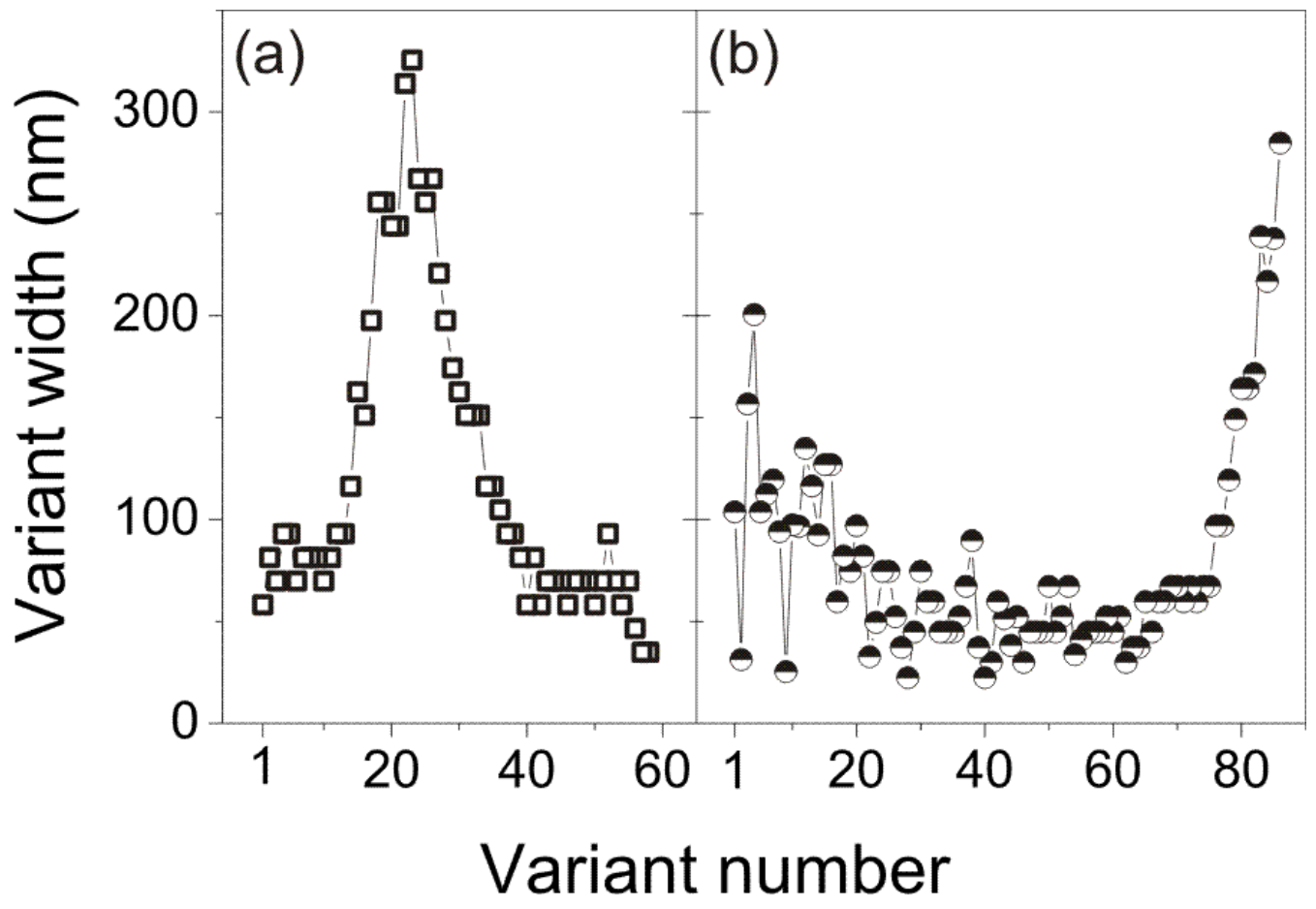

Figure 5: Quantitative analysis of variant width was performed for two different regions of type Y pattern. The variant numbers increase from the lower left corner to the upper right corner of the SEM micrographs in Figure 4. (a) The variant width of a long laminate (Figure 4a) shows a maximum in the middle of the laminate. (b) The variant width in a region between two long laminates (Figure 4d) does not show a clear trend.

Exemplarily, we show the analysis of variant width within two different regions of a type Y pattern. The first region represents a laminate with a length of about 200 µm (Figure 4a-c). The second region can be found in between four crossing long laminates (Figure 4d, upper/lower left and right corners). The length of the variants in between these crossing laminates does not exceed 2 µm (Figure 4d-f).

The distance between two intensity maxima in Hough space equals the distance between two twin boundaries (Figure 4c & f) and thus represents the variant width. The quantitative analysis of the variant widths from the different regions with type Y pattern (Figure 4) is shown in dependence of the variant number (Figure 5). The variant number is an arbitrary number that corresponds to the position of the variant within the SEM micrographs (Figure 4a & d). In our case, the variant number one was assigned to the smallest, still resolvable variant in the lower left corner of the SEM micrographs. Variants with higher numbers are situated in the upper right corners. The quantitative analysis of the variant width for the two different regions of type Y pattern (observed in Figure 4a & d) are shown in Figure 5a and b, respectively. In case of a long laminate (Figure 4a) the variant width continually increases from 58 nm up to a maximum of 325 nm with increasing variant number. With further increase of the variant number, the variant width decreases continuously again and remains constant in the range of 60 nm. The maximum variant width represent the variants situated in the middle of a long laminate (Figure 4a). It is interesting to compare this decrease of variant width with the fractal process of coarsening. It has recently been suggested, that the intermartensitic transition between 14M and NM proceeds in discrete steps, by doubling the variant width [17]. The continuous decrease of variant width within type Y pattern suggests that the present microstructure has a different origin than coarsening of NM martensite.

The variant width in the second region of type Y pattern (Figure 4d) does not show a clear trend. Up to the variant number of 18, the variant width ranges between 25 nm and 200 nm. This strong fluctuation can be caused by the refinement of variants when approaching the interface to a differently oriented type Y laminate (Figure 4d, lower left corner). The difference between the minimum and maximum values decreases with increasing variant number, the variant width oscillates around a mean value of 53 nm. Starting at variant number 76, the variant width increases continuously. This indicates the transition from intermediate type Y pattern to a long laminate (Figure 4a, Figure 5a).

We further analyzed the variant width of type X pattern by means of Hough transform (not shown here). The variant width is 209 nm in average with a standard deviation of 28 nm. In order to compare the results from Hough transform with the results we obtained by analyzing the AFM height profiles, the twinning periodicity has to be calculated. The twinning periodicity represents the sum of the widths of two adjacent variants. The value we determined by Hough transform is $(426 \pm 42)$ nm which is in good agreement with the twinning periodicity obtained by AFM measurements of $(412 \pm 73)$ nm [16].

**Structural analysis**

In order to probe the film structure, we performed X-ray diffraction measurements. The structural characterization by Θ-2Θ-measurements in Bragg-Brentano geometry is not sufficient for probing martensitic variants since the austenite-martensite transition is accompanied with a slight tilt and rotation of martensitic variants with respect to the substrate surface [18]. The Bragg-Brentano geometry only allows detecting lattice planes which are parallel to the substrate surface. Thus, the 002 peak of the epitaxial Cr buffer and the 004 peak of the austenitic Ni-Mn-Ga can be observed. In order to probe the martensitic structure of thin Ni-Mn-Ga films, we performed Θ-2Θ measurements in a tilted set-up with ω-offsets of up to 10°. The scans for each individual ω-offset for type X (Figure 6a) and type Y (Figure 6b) pattern are shown in an angular range between $60° < 2\Theta < 90°$ that corresponds to the range of 400 martensite peaks. When tilting the film from the film normal, we observe a large number of peaks that seem to merge. This has already been observed by Luo et al. [19] who explained the number of peaks with superstructure reflections of modulated martensite. Comparing the XRD scans of type X and type Y pattern for individual ω-offsets, differences in peak position as well as peak intensity are visible. This suggests that the underlying martensitic structure of type X and type Y are strongly different. In order to obtain a better comparison, we summed up the intensities of the individual scans for type X and type Y, respectively (Figure 6c). Despite the differences in the individual scans, the summed up intensity pattern show identical peak positions for type X and type Y. The summed up scans only vary in peak intensity and in the intensity ratio between different peaks, respectively. One can therefore conclude that type X and type Y pattern result from twinned laminates of the same martensitic phase but different orientations of the martensitic variants. This conclusion was confirmed by additional TEM investigations using Selected Area Diffraction (not shown here). These investigations revealed that both, type X and type Y pattern, result from twinning of modulated 14M martensite. However, the orientations of the underlying martensitic unit cells are different. While in case of type X the middle $b_{14M}$-axis lies within the film plane, it shows an out-of-plane direction for type Y. More detailed TEM investigations of the twin boundaries that create type X and type Y pattern and the underlying martensitic structure will be published separately.

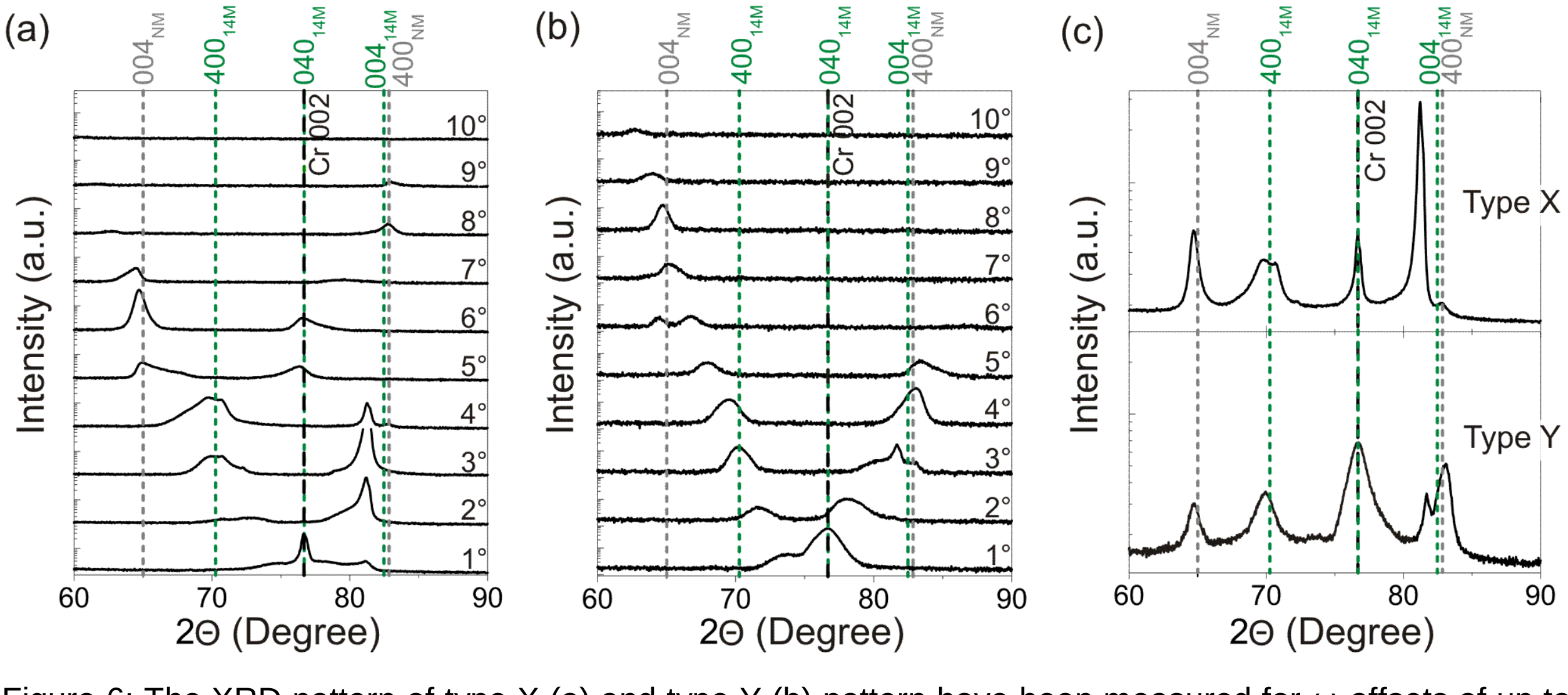

Figure 6: The XRD pattern of type X (a) and type Y (b) pattern have been measured for ω-offsets of up to 10° and suggest differences in the underlying martensite structure due to different peak positions and intensities. The summed up intensities (c) reveal identical peak positions for type X and type Y pattern, they only differ in peak intensity. The peak positions marked in the diffraction pattern are taken from Pons et al. [20].

## Discussion

Seiner et al. [10] recently proposed a scenario for the formation and growth of martensitic nuclei in an austenitic matrix. In this scenario, the transformation from austenite to martensite has been divided into four stages. It starts with the formation of single diamond nucleus, it proceeds with auto nucleation of neighboring diamonds, the transformation from diamond shape to parallelogram and end with the coalescence of individual diamonds and parallelograms, respectively. This four-stage formation mechanism can be used to explain the different microstructural pattern we observe on the surface of thin Ni-Mn-Ga films.

The orientation of the twin boundaries that form the respective pattern is the key difference between type X and type Y pattern. The twin boundaries are either inclined by 45° with respect to the film surface (type X pattern) or perpendicular to the film surface (type Y pattern). The origin of these different twin boundary orientations can be traced back to the orientation of the habit plane between austenite and martensite that is close to a $(101)_A$ plane of austenite [18]. There are six possible orientations of $(101)_A$ habit planes in a cubic crystal. In bulk Ni-Mn-Ga, all habit plane orientations are equivalent to each other. However, in thin Ni-Mn-Ga films one has to distinguish between two different orientations of habit planes due to the boundary conditions at the film-substrate-interface (Figure 1). There are four equivalent orientations of $(101)_A$ planes that are inclined by 45° with respect to the film surface and additionally there are two equivalent $(101)_A$ planes that are aligned perpendicular to the film surface (sketched in Figure 1). Accordingly, there are two different orientations for the martensitic nuclei that result in significantly different martensitic microstructures.

The martensitic nucleus itself contains eight habit variants that form a diamond shape. The long midribs of the diamond represent type I twin boundary whereas the shortest midrib is a compound twin boundary [10]. The growth of the diamond along the habit plane is not restricted. Therefore in bulk, diamond nuclei can exhibit an infinite length regarding the direction of growth along the habit planes. In case of thin films however, this length is restricted by the interface to the substrate or the film edges, respectively. The restrictive interface depends on the orientation of the habit planes. The transformation of a diamond nucleus to a parallelogram is realized by introducing a type II twin boundary between two habit plane variants that are not connected by a type I twin boundary in the initial state [10].

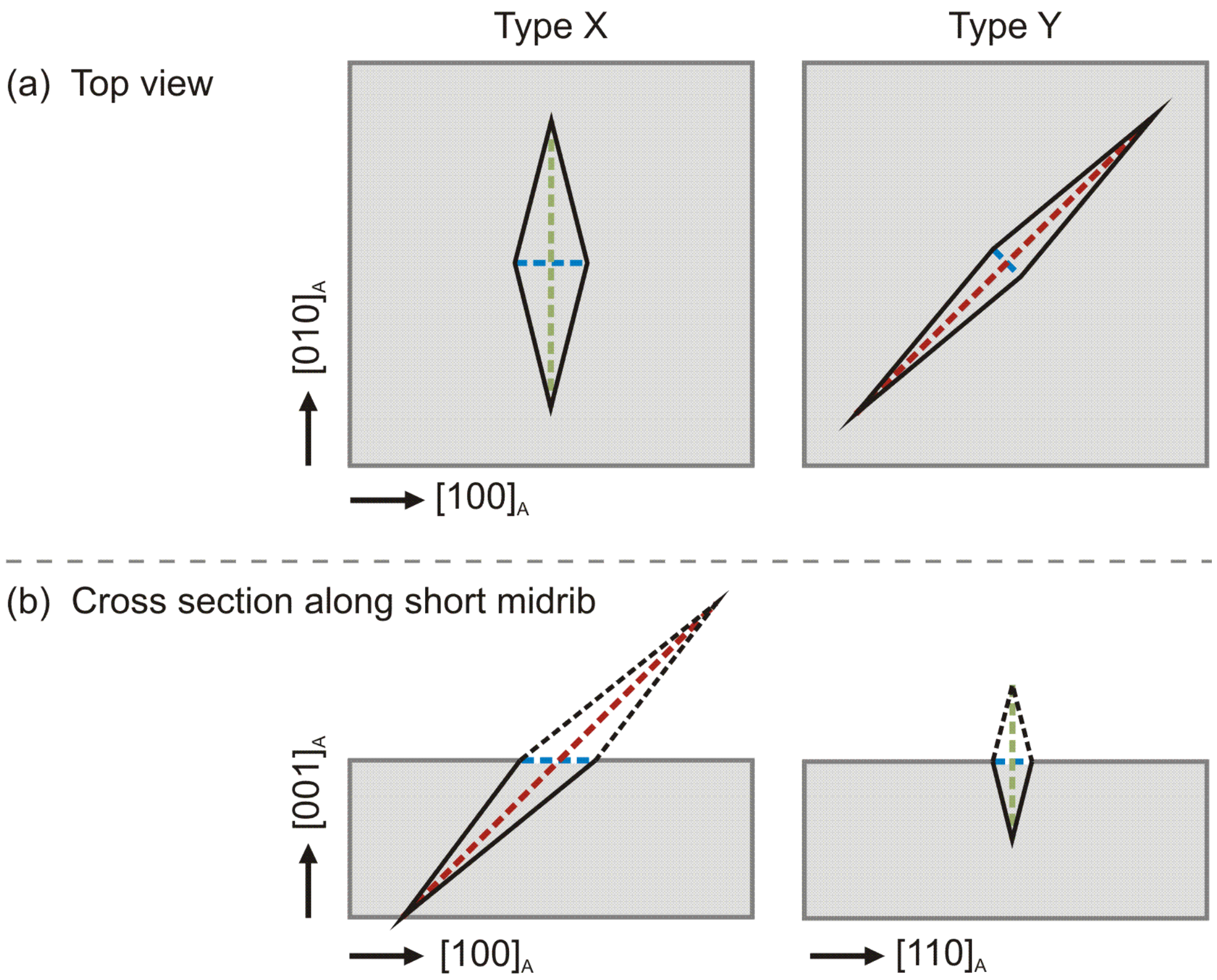

Figure 7: The diamond nucleus consists of eight habit plane variants with a short (blue), middle (green) and long (blue) midrib. (a) The top view shows the orientation of the diamond nucleus for type X and type Y pattern on the film surface. The cross cut along the short midrib of the diamond (blue line in (a)) shows that the long and middle midribs are either inclined by 45° (type X) or parallel (type Y) to the film normal $[001]_A$, respectively.

The orientations of the diamond nuclei for type X and type Y pattern are sketched in Figure 7. The nucleus contains eight habit plane variants which are connected by a short (blue line), middle (green line) and long (red line) midrib. Depending on the type of microstructural pattern, one can observe different midribs when looking at the film surface (top view in Figure 7a) or the cross section cut along the shortest midrib (Figure 7b).

In case of a type X pattern, the diamond shaped nucleus is tilted from the film surface by an angle of 45°. This is sketched in Figure 7b. The long midrib which is referred to as the length of the diamond (red line) can grow until it meets the incompatible interface to the substrate. The middle and short midribs may slightly continue to grow but their lengths will be limited by a balance of the incompatible interface area and the volume energy driving the transformation. The observed surface pattern suggests, that instead of growing further, the diamond nucleus rapidly elongates into a parallelogram. This elongation is accompanied by the formation and growth of a type II twin boundary that represents the midrib of the parallelogram. The length of the parallelogram and thus the length of the type II twin boundary can increase further until it crosses another individual parallelogram. The proposed nucleation scenario explains the similar width of the martensitic variants observed at the sample surface of type X pattern [15] and gives an alternate explanation to their thickness dependency. Since the variant width does not vary throughout the surface of the respective Ni-Mn-Ga films, we can conclude that the proposed auto nucleation scenario [10] does not play an important role in case of type X pattern.

The microstructural pattern of type X shows further consequences of the nucleation scenario. The traces of twin boundaries visible at the film surface (Figure 2 a) deviate from the $[010]_A$ direction of austenite by about ±6°. This positive or negative angular deviation is expected for a type II twin boundary that represents the midrib of the parallelogram. The angle of about 12° between twin laminates was observed independently in SEM micrographs (Figure 2 a) and in integral pole figure measurement of the $004_{14M}$ reflection of 14M martensite (not shown here). The observed angle of 12° is in good agreement with the projected deviation of 12.2° that we calculated for a mesoscopic type II twin boundary between $a_{14M}$-$b_{14M}$-laminates with an underlying $c_{NM}/a_{NM}$-ration of 1.20. The angular deviation originates from the boundary conditions at the rigid substrate. In order to keep coherence with the substrate, type II twin boundaries have to alternate between positive and negative deviations in thin Ni-Mn-Ga films. This alternation results in a rhombus like pattern observed at the substrate surface (Figure 2a).

The TEM cross section of type X pattern (Figure 2 b) shows that the twin width decreases when approaching the interface between film and the rigid substrate. Previously, this refinement of the twin microstructure has been explained with gaining elastic energy at the interface on the cost of twin boundary energy [21]. The nucleation scenario suggests an alternate explanation. Although the wedge of a parallelogram is quite sharp (3°) [10] its sides are not exactly parallel. Therefore neighboring parallelograms form a small gap in between them that enclose residual austenite. This gap is closed by sharp needles that presumably form during further undercooling in order to transform residual austenite to martensite.

In case of the observed type Y pattern, the $(101)_A$ habit plane is almost perpendicular to the film surface (Figure 7). The transformation from austenite to martensite only requires forming one half of a diamond (wedge) at the film surface (Figure 7). In contrast to the habit plane variants that result in type X pattern, the length of the diamond (red line in Figure 7a) is only restricted by the films edges or the crossing with another individual diamond. For that reason, the twin laminates can show lengths of up to several hundreds of micrometer. These long laminates can be described by a super diamond which is formed by auto nucleation of neighboring diamond nuclei [10]. This concept of auto nucleation can be supported by the observed trend in variant width (Figure 5a) within one region of type Y pattern (Figure 4a). The variant width shows a maximum in the middle of such a long laminate, it decreases continuously on both sides of this maximum value. However, the value of twinning periodicity is more significant, since a diamond nucleus is composed of two adjacent variants. Since the variants that form a diamond cannot be determined unambiguously from SEM micrographs, the clear trend in variant width is used to emphasize the auto nucleation scenario in this case.

Since the midribs of the diamonds represent type I twin boundaries, the majority of the twin boundaries that form the type Y pattern can be assigned to type I twin boundaries. The expansion from diamond to parallelogram could only be observed in individual cases of twin boundaries forming a type Y pattern.

The nucleation scenarios sketched above explain the differences in the surface pattern we observed in experiments. However, we have not been able to clearly identify the driving force for the formation of twin boundaries that are differently aligned with respect to the substrate normal and thus the driving force for the formation of either type X or type Y pattern at the film up to now. We can however exclude the influence of single parameters such as deposition temperature, pressure or film composition.

It was suggested that differences internal stress of the thin film can influence the formation of the different microstructures, since twin boundaries can accommodate external stresses in a different way depending on their orientation with respect to the substrate surface [22]. In this case, a biaxial stress state would favor the formation of twin boundaries that create a type X pattern since the nucleation of a complete diamond is necessary. Furthermore, there are four equivalent orientations of tilted $(101)_A$ twin boundaries that form the type X pattern. As a consequence, the formation of type Y pattern which contains twin boundaries that are perpendicular to the substrate surface should be favored by a stress free state. In this case, there are only two possible orientations and the martensitic nucleus only consists of one half of the diamond. The internal stress state of a thin film can be influenced by multiple parameters such as film thickness, impurities or other lattice defects.

## Conclusion

The analysis of Ni-Mn-Ga films with different types of microstructural pattern at the film surface revealed, that both type X and type Y pattern are formed by mesoscopic twin boundaries of 14M martensite. The only difference between type X and type Y pattern lies in the orientation of the martensitic unit cell. Consequently, the orientation of 14M twin boundaries with respect to the film surface differs between type X and type Y pattern. Whereas twin boundaries that form a type X pattern are tilted from the film surface by about 45°, type Y pattern is governed by twin boundaries aligned perpendicular to the film surface. These different orientations originate from the orientations of the martensitic nuclei which are equally tilted from the film surface or perpendicular to it. The resulting twin boundaries represent either type II or type I twin boundaries, respectively, with significant differences in twinning stress [23]. Although, we did comprehensive analyses of several influential parameters, we could not detect one single parameter which might be responsible for the formation of the differently oriented martensitic nuclei. It rather seems as if the interaction of multiple parameters plays a large role in the formation of the martensitic nuclei and their growth. Furthermore we have to take the influence of internal properties such as internal mechanical stress into consideration.

We can draw conclusions from the type of microstructural pattern visible at the film surface on the applicability of the respective Ni-Mn-Ga film as active material. Type X pattern mostly consist of type II twin boundaries, whereas type Y microstructure is governed by type I twin boundaries. The twinning stress of type II twin boundaries is about an order of magnitude lower than the value for type I twin boundaries [23]. Therefore, Ni-Mn-Ga films with a high fraction of type X pattern should be favorable regarding the application potential as active material. The type II twin boundaries forming this pattern can be moved with lower values of external mechanical stress and therefore in lower magnetic fields.

**Acknowledgement**

The authors would like to thank the DFG for funding this work via the priority program SPP1239 and SPP1599.